# Optical Properties of Pure and Mixed Germanium and Silicon Quantum Dots

[1]Shanawer Niaz[*], [2]Aristides D. Zdetsis, [1]Manzoor Ahmad Badar, [1]Safdar Hussain, [3]Imran Sadiq, [4]Muhammad Aslam Khan

[1] Department of Physics, University of Sargodha, Sargodha, PK-40100, Pakistan

[2] Molecular Engineering Laboratory, Department of Physics, University of Patras, Patras, GR-26500, Greece

[3] Centre of Excellence in Solid State Physics, University of the Punjab, Lahore, Pakistan

[4] Department of Physics, BUITEMS, Quetta, 87300, Pakistan

shanawersi@gmail.com, shanawersi@uos.edu.pk [*]

**Summary**

We study the optical properties of hydrogen passivated silicon, germanium and mixed Ge/Si core/shell quantum dots (QDs) using high accuracy Density Functional Theory (DFT) and time-dependent DFT (TDFT). We employ the hybrid DFT functional of Becke, Lee, Yang and Parr (B3LYP) in combination with good quality basis sets. As we have shown in our previous work, this combination is an accurate and computationally efficient way for such calculations. The mixed quantum dots, as would be expected, are more versatile and offer more possibilities for band gap engineering, with gap values (electronic and optical) between those of the corresponding Si and Ge dots. Our results support the quantum confinement theory for all three types of QDs.

**Introduction**

Similarly to pure silicon, pure germanium as well as mixed Ge/Si quantum dots (QDs) has been a favourite subject to study visible photoluminescence (PL) [1-10]. Most of the work has been done to correlate the optical and electronic characteristics to the diameter of these materials, mostly for "band gap engineering". Quantum confinement (QC) model has been widely accepted in dealing with the band-gap versus size correlation. This is true for both Si [4-6] and Ge [7-8] QDs. Quantum confinement is responsible for the widening of the gap of Si QDs from the bulk value of 1.1 eV to values of 2-3 eV (for larger nanocrystals) up to 6-7 eV for smaller nanocrystals [4-5]. This widening of the gap for small size quantum dots obviously is equally undesirable as the shrinking of the gap (in bulk Si). This problem at small sizes of nanocrystals is not so strong for Ge nanocrystals [7] compared to Si because of the smaller band gap of bulk Ge. Even in this case, the possibilities of adjusting the optical gap (and the band gap) are limited only to proper size selection. The possibility of combining the advantages of Si not only in the optical, but also in the electronic and mechanical properties with those of Ge is an exciting and subject with great potential. It is unnecessary to say, that, not only the minimum critical diameter of the nanocrystals for visible photoluminescence is important, but also the maximum possible diameter. With this in mind, we have examined the optical and electronic properties of mixed nanocrystals of the form Core/Shell GeSi which is most preferable when Ge is in core [9].

In our present work we study structural, electronic and optical properties of pure silicon and germanium quantum dots as well as examine ways for optimal "band gap engineering" (band gap adjustment to desired values) of SiGe quantum dots. Both, Si and Ge, as well as Si/Ge core/shell nanocrystals follow the principle(s) of Quantum confinement.

**Technical details**

In this work we present *ab initio* calculation of structural optimization and electronic properties using density functional theory while the time dependent and excited state properties (optical gap) of Si, Ge and Si/Ge nanocrystals based on time dependent density functional theory (TDDFT) [11] employing the hybrid nonlocal exchange-correlation functional of Becke and Lee, Yang and Parr (B3LYP) [12]. The B3LYP functional has enough evidence that it can efficiently reproduce the band structure of a wide variety of materials, including crystalline Si and Ge [13]. In our present work the quantum dots ranges from 17 to 147 atoms, with 36 to 100 H atoms (a total of about 247 atoms) in each case. The diameter of the largest clusters used in this study is 18.36 Å for Si, 19.51Å for Ge and 19.28 Å for mixed one. All dots carry $T_d$ symmetry, their geometries have been fully optimized within this symmetry constrain.

The DFT and the TDDFT calculations were performed with the TURBOMOLE [14] suite of programs using orbital basis sets of split valence [SVP]: [4s3p1d] [15] in the case of Si. In case of Ge quantum dots, we apply a suitable effective core potential (ECP) to reduce the computation cost and to extend to as large QDs as possible, without sacrificing the quality and accuracy of the results. We found, by comparing to similar all electron calculations (taking into account for structural and electronic properties), that the choice of LANL2DZ ECPs combined with LANL08d basis set [3s4p1d] is one of the best possible combinations [16-18].

**Results and discussion**

Fig. 1 shows some of the optimized structures of silicon, germanium and mixed quantum dots. The geometry optimization was performed using DFT/B3LYP for all dots. Fig. 2(a) shows the silicon-silicon bond length distribution diagram for pure silicon quantum dots. Obviously, bond lengths have been calculated after optimization of the structures. The average Si-Si bond length is 2.38 Å; whereas the Ge-Ge bond length distribution diagram in Fig. 2(b) shows an average bond length of 2.48 Å. Further structural properties (diameters and composition etc) are given in Table-1.

We have constructed mixed Ge/Si core/shell quantum dots in the following compositions: 30/5, 30/17, 54/17, 70/29, and 88/59 to complete 35, 47, 71, 99 and 147 total number of atoms (excluding hydrogen atoms) in order to compare with the corresponding pure Si or Ge clusters. We have also examined a limited number of core/shell systems with incomplete shells, in which the number of shell Si atoms is smaller than the number required for a complete filling of the shell, and actually smaller than the number of core Ge atoms. In such cases, denoted as Ge/Si, in which the concentration of germanium in the core is higher than the concentration of silicon in the (incomplete) shell, the reverse ratio of atoms was used: 29/6, 29/18, 53/18, 71/28, and 87/60. The diameter of largest Si/Ge and Ge/Si dots was about 18.8 Å (18.79 Å for Si/Ge and 18.82 Å for Ge/Si)

As we can see in Fig. 3(a), the difference between the optical and HOMO-LUMO gaps for Si dots is of the order of 0.5 eV, except for the very small (strongly confined) $Si_{17}H_{36}$ dot, in which the difference is about 1eV. The corresponding difference for Ge in Fig. 3(b) is about 0.4 eV (except for the smaller dots). The small fluctuations in the differences are apparently due to the different influence of quantum confinement in HL and optical gaps for different sizes. Besides the many body effects, the optical gap is also influenced by the corresponding selection rules.

The size dependence of band gap of the nanoparticles according to quantum confinement model is expected to be of the form:

$$E_g = A + B * D^{-n} \qquad (1)$$

where A, B, D and n are adjustable parameters to be determined by the fit. It is clear (if we consider the limit D→∞) that the value of the parameter A corresponds to the band gap of the infinite Si or Ge crystal. Different values of the parameter n exist in the literature however, as we show in the previous study [19] for Si quantum dots. A distinct value of $n = 1$ is more appropriate, if diameter is concerned. The fitted curves for all structures follow the dependence described by the QC equation 1, the quality of the fit can be observed in Fig. 3 whereas parameter values can be found in equation 2 for silicon and in equation 3 for germanium respectively.

$$E_{g,HL} = (1.46 \pm 0.24) + (40.24 \pm 2.97) \times D^{-1}$$

$$E_{g,OPT} = (1.16 \pm 0.20) + (39.60 \pm 2.54) \times D^{-1}$$
(2)

$$E_{g,HL} = (1.21 \pm 0.17) + (43.18 \pm 2.25) \times D^{-1}$$

$$E_{g,OPT} = (0.95 \pm 0.17) + (39.27 \pm 2.27) \times D^{-1}$$
(3)

In this graphs of Fig. 3(a and b), we have used only a limited number of data points neglecting large size dots, the presence of which is crucial for the band gap value at infinity (employing equation 1). Therefore we do not expect to have good value of the infinite gap. Nevertheless, if we consider the maximum uncertainty of the parameter A ($\pm 0.20\ eV$), for the optical gap, the band gap value is not so bad. This is also true for the HL gap. It is clear that if we enrich our data points with results from large quantum dots we would recover the correct band gaps.

Concerning germanium, we can clearly not go to much larger sizes due to the size limitations of all-electron calculations and to the limitation of the ECP accuracy for such type of calculations. Nevertheless, the optical gap is obtained with surprisingly good accuracy ($0.95 \pm 0.17\ eV$), although this is not quite true for the HL gap. Also, as we can see in Fig. 3(c), which shows the gap variation of the mixed Si/Ge dot, although we would have expected the gaps to be between those of the pure elemental nanocrystals, this is not quite so. Apparently, on top of possible uncertainties due to poorer description of Ge atoms, this is partly due to incomplete shell nanocrystals.

Besides, the structural, electronic and optical characteristics, the cohesive properties (binding and/or cohesive energy, and "interaction energy per heavy atom", as will be defined below) are of primary importance. It has been demonstrated recently by Zdetsis et al. [20, 21] that the size variation of nanocrystals (in particular of $BeH_2$, and $MgH_2$) can give valuable and accurate information about the cohesive properties (binding energy, desorption energy, etc.) of the corresponding infinite crystal. Therefore, it would be very important to examine the size variation of the cohesive energy of the Si dots, and the possibility of evaluating the cohesive energy of the infinite crystal. To this end, in Fig. 4 we have plotted the binding energy per heavy atom dependence on the number of heavy atoms, N. It is clear from this figure that pure Si dots show high binding energy per silicon atom compared to the pure Ge dots as would be expected on the basis of the relative bond energy, whereas mixed dots are somewhere in between. For instance, the average binding energy difference between silicon and germanium dots is about 0.87 eV/atom (see figure 4 and Table-1). In contrast, one can predict that high concentration of Si atoms, in mixed quantum dots, increases the binding energy values. Furthermore, the energy dips shown in Fig. 4 for N=29, 35 are possibly, due to the high ratio of number-of-heavy-atoms to the number-of-hydrogen-atoms (on surface) compared to the same ratio for N=17 and N=47.

To study the core shell interaction we have constructed Ge-Si quantum dots as germanium in core and silicon on shell with constant number of core atoms and varying number of shell atoms. For instance in the range of 17 to 147 total number of atoms, we fix 5 germanium atoms in core and vary silicon on shell as 12, 24, 30, 42, 66, 94 and 142 to complete the total number of 17, 29, 35, 47, 71, 99 and 147 atoms to compare properties with the dots we already have discussed for pure silicon and germanium quantum dots. We consider the same technique to develop in extension to these quantum dots as 29, 35, 47, and 71 germanium atoms in core as well.

We define the vertical core/shell interaction energy per heavy atom, $\Delta E_{Ge_mSi_nH_l}$, of the $Ge_mSi_nH_l$ core / shell nanocrystal, consisting of a $Ge_m$ core (here m=5, 17) and a $Si_nH_l$ shell of $n$ silicon atoms and $l$ passivated hydrogen atoms, by the relation:

$$\Delta E_{Ge_mSi_nH_l} = \frac{1}{(n+m)}\{E[Ge_m]_0 + E[Si_nH_l]_0 - E_{tot}[Ge_mSi_nH_l]\} \qquad (4)$$

where $E[Ge_m]_0$ and $E[Si_nH_l]_0$ are the total energies of the separate individual core and shell un-relaxed structures, respectively. This energy is a measure of the Si/Ge core-surface interaction.

As we can see in the diagrams of Fig. 5 (and Table-2), the core/shell interaction energy per heavy atom increases with the increase of Ge core size, apparently due to the increasing number of the surface Ge atoms (and, therefore, increasing number of the surface Si – Ge pairs). For example, the interaction energy per heavy atom is about three times higher for m=17 compared to m=5. As far as the total number of heavy atoms (Si + Ge) increases the interaction energy per heavy atom decreases since the number of interacting surface atoms is constant. Eventually, the interaction energy per heavy atom is expected to saturate to very small number close to zero for infinite number of shell atoms (where core atoms are constant).

**Conclusion**

Within limited size range (20 Å) of the pure silicon and germanium quantum dots, the predicted HOMO-LUMO gap values are higher than expectations (more explicit for germanium). Nevertheless, the optical gap for both quantum dots is obtained with surprisingly good accuracy compared with HL gap. Apparently, more data points are required for improved band gap value at infinity, which is quite hard for germanium due to the size limitations of all-electron calculations and to the limitation of the ECP accuracy. Binding energy plots versus the number of heavy atoms confirms that pure silicon has higher binding energy compared to pure germanium, which is rather obvious. Concerning mixed Si-Ge dots, the peculiar HOMO-LUMO gap behavior with respect to their size reflects the way they have constructed, and obviously the extrapolation to infinity has no meaning. Clearly, the core/shell interaction energy per heavy atom increases with the increase of Ge core size, whereas, when total number of heavy atoms (Si + Ge) increases the interaction energy per heavy atom decreases since the number of interacting surface atoms is constant.

**Acknowledgments**

This project was funded by Higher Education Commission of Pakistan (Project No. 72). Computational resources support from the National Centre for Physics (NCP) Pakistan is also gratefully acknowledged.

Table-1: Structural (total number of atoms, symmetry and diameter in Angstroms), Energetic (binding energy), electronic (HOMO-LUMO gaps) and selected optical (optical gaps) characteristics of silicon and germanium and mixed quantum dots respectively.

| Total # atoms | Sym | Silicon | | | | | Germanium | | | | |
|---|---|---|---|---|---|---|---|---|---|---|---|
| | | Cluster | D (Å) | H-L Gap (eV) | Opt. Gap (eV) | BE (eV/atom) | Cluster | D (Å) | H-L Gap (eV) | Opt. Gap (eV) | BE (eV/atom) |
| 53 | $T_d$ | $Si_{17}H_{36}$ | 9.62 | 5.72 | 5.03 | 9.21 | $Ge_{17}H_{36}$ | 10.62 | 5.34 | 4.70 | 8.27 |
| 65 | $T_d$ | $Si_{29}H_{36}$ | 10.44 | 5.14 | 4.52 | 7.09 | $Ge_{29}H_{36}$ | 10.87 | 5.08 | 4.47 | 6.21 |
| 71 | $T_d$ | $Si_{35}H_{36}$ | 11.46 | 5.01 | 4.39 | 6.60 | $Ge_{35}H_{36}$ | 11.94 | 4.93 | 4.34 | 5.74 |
| 107 | $T_d$ | $Si_{47}H_{60}$ | 13.45 | 4.62 | 4.02 | 7.20 | $Ge_{47}H_{60}$ | 14.02 | 4.26 | 3.70 | 6.33 |
| 155 | $T_d$ | $Si_{71}H_{84}$ | 14.24 | 4.15 | 3.59 | 6.95 | $Ge_{71}H_{84}$ | 14.57 | 4.06 | 3.53 | 6.06 |
| 199 | $T_d$ | $Si_{99}H_{100}$ | 16.62 | 3.93 | 3.40 | 6.55 | $Ge_{99}H_{100}$ | 17.39 | 3.74 | 3.26 | 5.70 |
| 247 | $T_d$ | $Si_{147}H_{100}$ | 18.36 | 3.60 | 3.12 | 5.77 | $Ge_{147}H_{100}$ | 18.91 | 3.50 | 3.05 | 4.94 |
| | | **Si/Ge Mixed** | | | | | **Ge/Si Mixed** | | | | |
| 71 | $T_d$ | $Ge_5Si_{30}H_{36}$ | 11.59 | 4.78 | 4.36 | 6.49 | $Ge_{29}Si_6H_{36}$ | 11.95 | 4.93 | - | 5.90 |
| 107 | $T_d$ | $Ge_{17}Si_{30}H_{60}$ | 13.7 | 4.52 | 3.91 | 6.93 | $Ge_{29}Si_{18}H_{60}$ | 13.84 | 4.40 | - | 6.71 |
| 155 | $T_d$ | $Ge_{17}Si_{54}H_{84}$ | 14.02 | 4.07 | - | 6.78 | $Ge_{53}Si_{18}H_{84}$ | 14.41 | 4.05 | - | 6.35 |
| 199 | $T_d$ | $Ge_{29}Si_{70}H_{100}$ | 16.98 | 3.80 | - | 6.34 | $Ge_{71}Si_{28}H_{100}$ | 17.37 | 3.70 | - | 5.98 |
| 247 | $T_d$ | $Ge_{59}Si_{88}H_{100}$ | 18.79 | 3.44 | - | 5.46 | $Ge_{87}Si_{60}H_{100}$ | 18.82 | 3.48 | - | 5.30 |

Table-2: Core energies, Shell energies and Total energies of Ge/Si (Core/Shell) quantum dots for two different Ge cores (m=5, 17).

| | 5 germanium atoms in core | | | | 17 germanium atoms in core | | | |
|---|---|---|---|---|---|---|---|---|
| Cluster | Total No. of Atoms | Core Energy (eV) | Shell Energy (keV) | Total Energy (keV) | Cluster | Total No. of Atoms | Core Energy (eV) | Shell Energy (keV) | Total Energy (keV) |
| $Ge_5Si_{24}H_{36}$ | 29 | -507.89 | -189.55 | -190.09 | $Ge_{17}Si_{12}H_{36}$ | 29 | -1733.11 | -95.01 | -96.83 |
| $Ge_5Si_{30}H_{36}$ | 35 | -507.89 | -236.80 | -237.33 | $Ge_{17}Si_{18}H_{36}$ | 35 | -1733.14 | -142.26 | -144.08 |
| $Ge_5Si_{42}H_{60}$ | 47 | -507.75 | -331.68 | -332.22 | $Ge_{17}Si_{30}H_{60}$ | 47 | -1732.92 | -237.15 | -238.96 |
| $Ge_5Si_{66}H_{84}$ | 71 | -507.89 | -521.06 | -521.59 | $Ge_{17}Si_{54}H_{84}$ | 71 | -1732.44 | -426.53 | -428.34 |
| $Ge_5Si_{94}H_{100}$ | 99 | -507.82 | -741.81 | -742.34 | $Ge_{17}Si_{82}H_{100}$ | 99 | -1733.07 | -647.28 | -649.09 |
| $Ge_5Si_{142}H_{100}$ | 147 | -507.94 | -1119.80 | -1120.33 | $Ge_{17}Si_{130}H_{100}$ | 147 | -1732.58 | -1025.28 | -1027.07 |

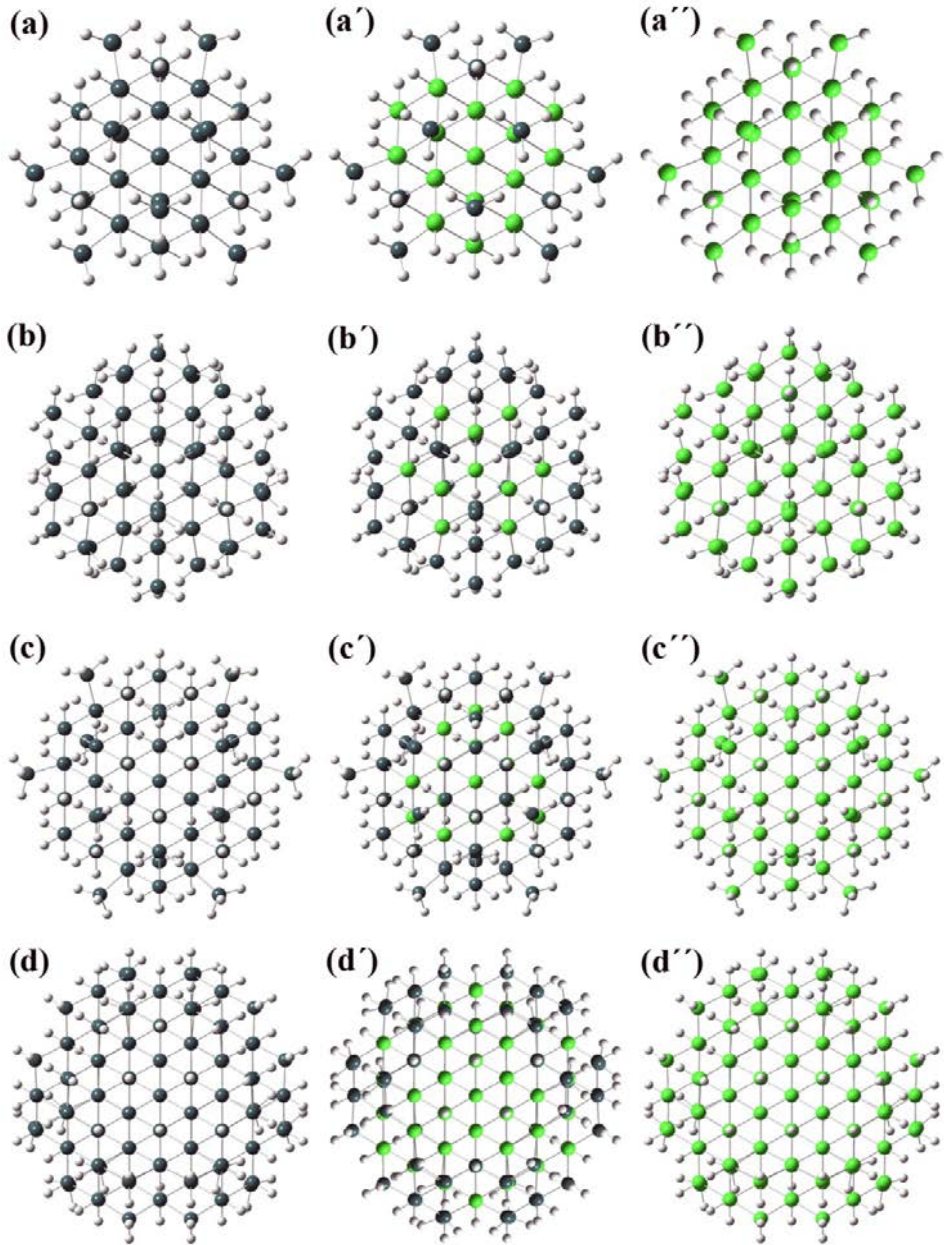

Fig. 1: Optimized structures of representative pure silicon (a, b, c, d); mixed silicon/ germanium (a′, b′ c′, d′); and pure germanium quantum dots (a′′, b′′, c′′, d′′) of various sizes and compositions. The compositions for pure silicon dots are: (a) $Si_{47}H_{60}$ (b) $Si_{71}H_{84}$ (c) $Si_{99}H_{100}$ (d) $Si_{147}H_{100}$. For mixed Si/Ge the structures (a′, b′, c′, d′) correspond to compositions $Ge_{29}Si_{18}H_{60}$, $Ge_{17}Si_{54}H_{84}$, $Ge_{29}Si_{70}H_{100}$, $Ge_{87}Si_{60}H_{100}$ respectively. The pure germanium dots (a′′, b′′, c′′, d′′) are fully analogous to the silicon dots (a, b, c, d). All structures were relaxed using DFT/B3LYP method.

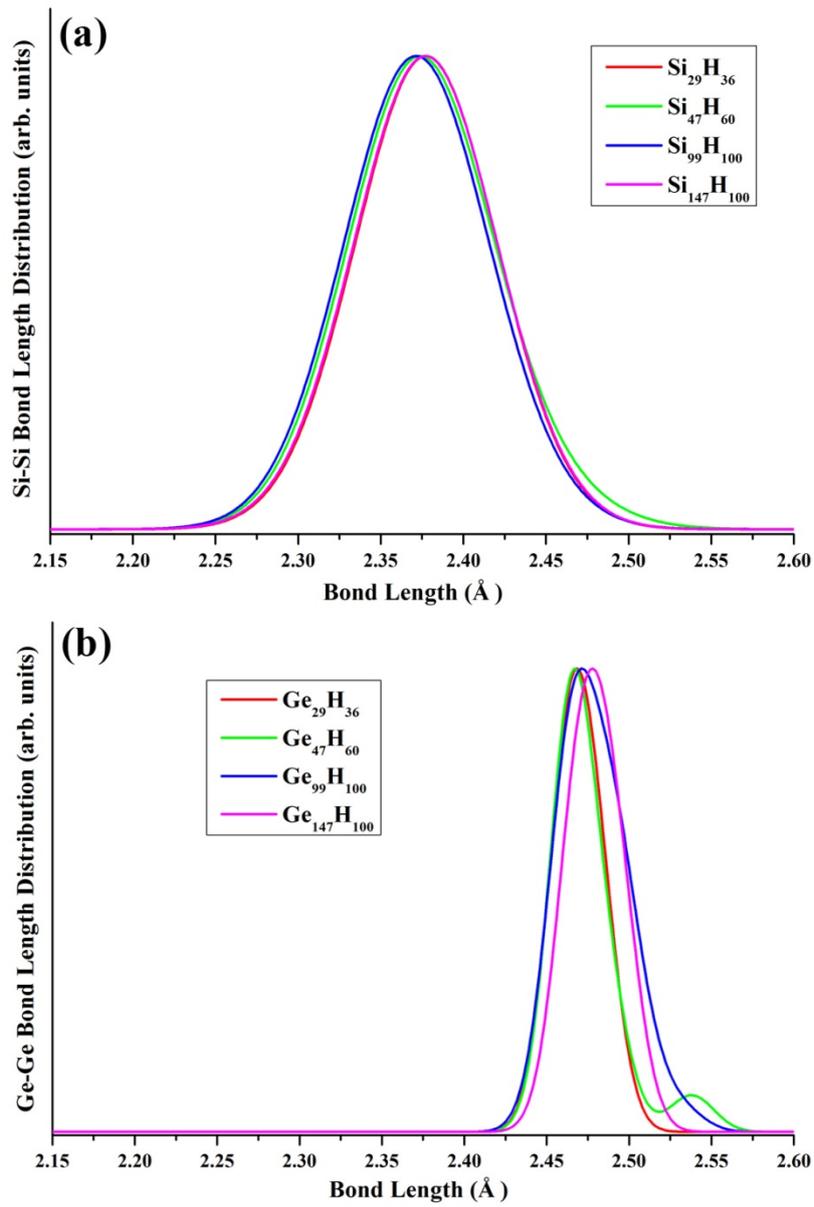

Fig. 2: (a) Silicon-silicon, and (b) germanium-germanium bond length distribution diagrams for pure silicon and germanium quantum dots, respectively.

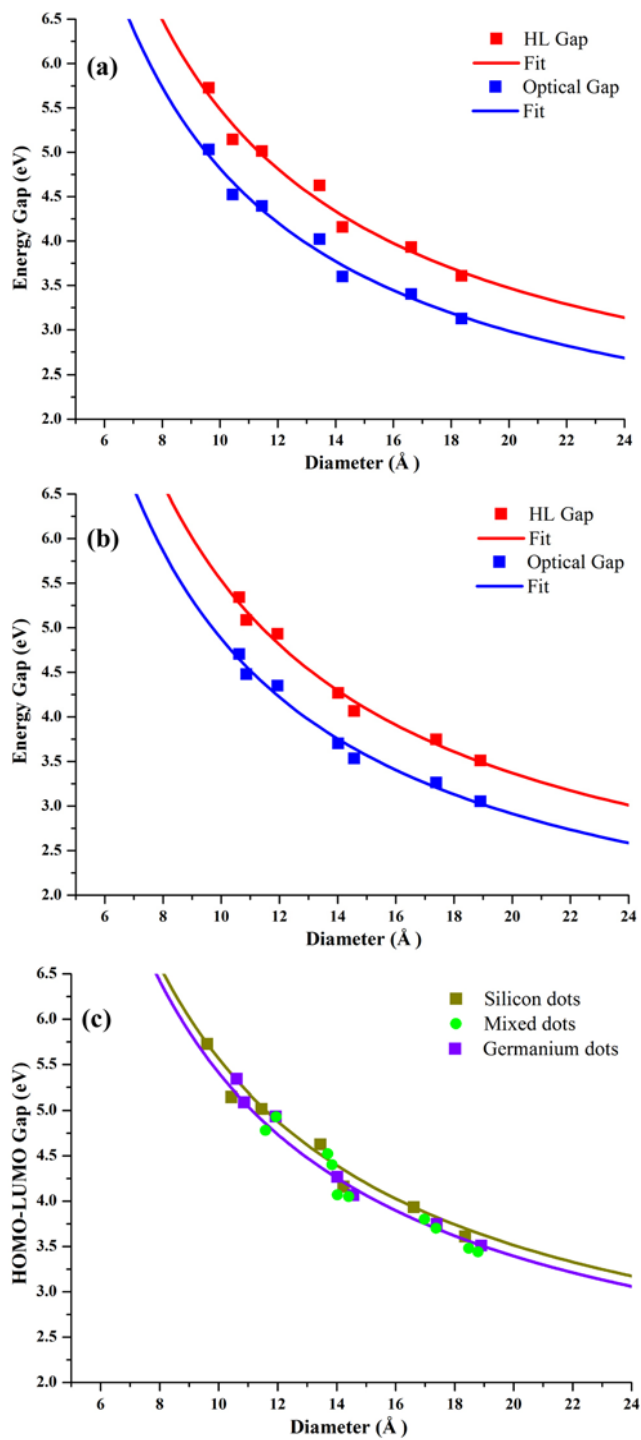

Fig. 3: Optical and HL gap (blue and red squares respectively) of Silicon (a), Germanium (b) quantum dots versus their size (diameter). Blue and red solid lines correspond to the usual "quantum confinement" fit described by equation 1, the parameters of which are shown in the figures (a, b). In Fig. 3(c) we show the D-dependence of HOMO-LUMO gap for mixed dots in comparison to their elemental Si and Ge dots.

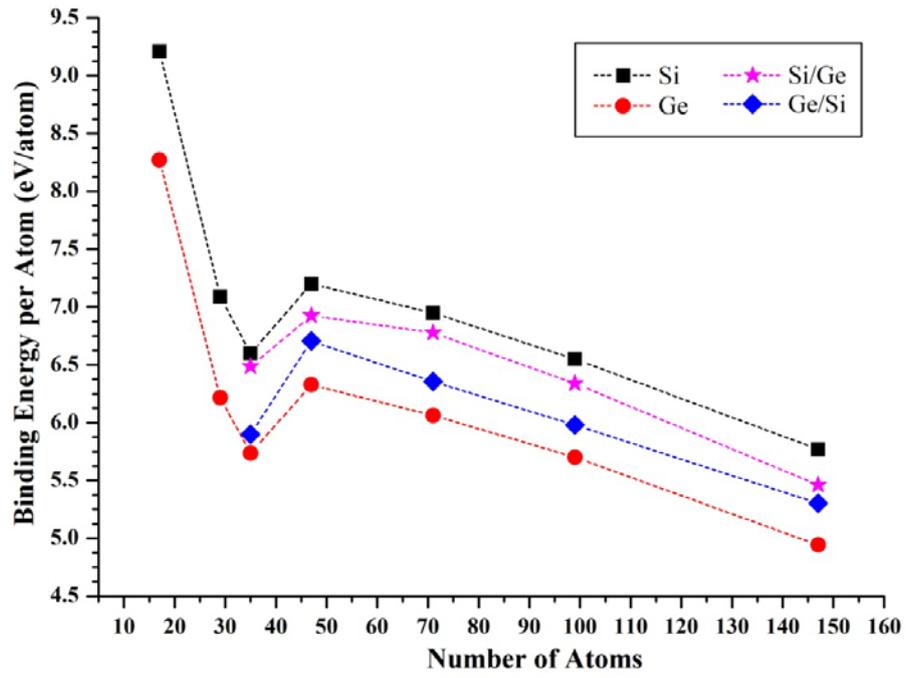

Fig. 4: Binding energy plot of Si, Ge, and their mixed quantum dots with respect to size of the dots (i.e. total number of heavy atoms).

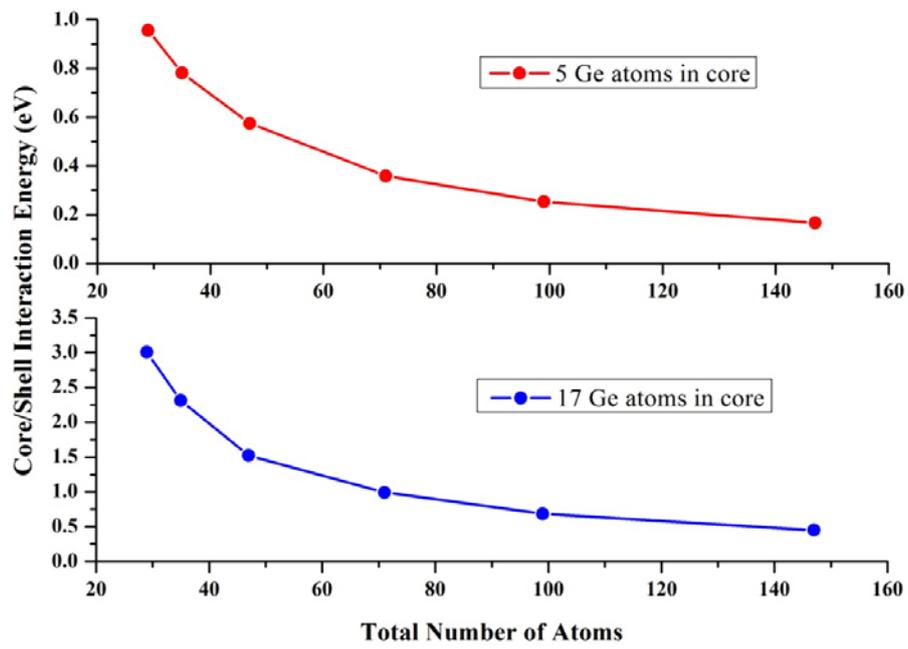

Fig. 5: Variation of the core/shell vertical interaction energy per heavy atom $\Delta E_{Ge_mSi_nH_l}$ with the total number of Si + Ge atoms, for two different Ge cores (m=5, 17).